\documentclass[12pt]{iopart}

\usepackage[ansinew]{inputenc}
\usepackage[english]{babel}
\usepackage{amssymb}
\usepackage[matrix,arrow,curve,frame]{xy}
\usepackage{url}
\usepackage{subeqn}

\newcommand{\mb}[1]{\mathbb{#1}}

\newcommand{\braket}[2]{\left \langle #1 \, \left | \right . \, #2 \right \rangle }
\newcommand{\braxket}[3]{\left \langle #1 \, \left | \, #2 \, \right | \, #3 \right \rangle }
\newcommand{\ket}[1]{\left | \, #1 \right \rangle}

\newcommand{\oneI}{1\hspace{-0.1cm}{\rm I}}

\newcommand{\modulo}[1]{\left | #1 \right |}

\newcommand{\fracd}[2]{{\displaystyle \frac{#1}{#2}}}

\newcommand{\urlbib}[1]{{\footnotesize{\url{#1}}}}

\newenvironment{ssfenumerate}
{\begin{enumerate}\normalfont\sffamily}
{\end{enumerate}}

\begin{document}
\renewcommand{\thefootnote}{\arabic{footnote}}

\title{Quantum hypercomputation based on the dynamical algebra $\mathfrak{su}(1,1)$}

\author{A. Sicard, J. Ospina, M. Vélez}

\address{Group of Logic and Computation, EAFIT University, AA. 3030, Medellín, Colombia}

\eads{\mailto{asicard@eafit.edu.co}, \mailto{judoan@epm.net.co}, \mailto{mvelez@eafit.edu.co}}

\begin{abstract}
An adaptation of Kieu's hypercomputational quantum algorithm (KHQA) is presented. The method that was used was to replace
the Weyl-Heisenberg algebra by other dynamical algebra of low dimension that admits infinite-dimensional irreducible representations with naturally defined generalized coherent states. We have selected the Lie algebra $\mathfrak{su}(1,1)$, due to that this algebra posses the necessary characteristics for to realize the hypercomputation and also due to that such algebra has been identified as the dynamical algebra associated to many relatively simple quantum systems. In addition to an algebraic adaptation of KHQA over the algebra $\mathfrak{su}(1,1)$, we presented an adaptations of KHQA over some concrete physical referents: the infinite square well, the infinite cylindrical well, the perturbed infinite cylindrical well, the Pöschl-Teller potentials, the Holstein-Primakoff system, and the Laguerre oscillator. We conclude that it is possible to have many physical systems within condensed matter and quantum optics on which it is possible to consider an implementation of KHQA.
\end{abstract}


\pacs{03.67.Lx}

\submitto{\JPA}

\maketitle

\renewcommand{\thefootnote}{\arabic{footnote}}
\section{Introduction}
The hypercomputers compute functions or numbers, or more
generally solve problems or carry out tasks, that cannot be
computed or solved by a Turing machine (TM)~\cite{Copeland-2002,Sicard-Velez-2001a-english}. Starting from
that seems to be the first published model of hypercomputation,
which is called the Turing's oracle machines \cite{Turing-1939};
the formulations of models and algorithms of hypercomputation
have applied a wide spectrum of underlying theories~\cite{Copeland-2002,Stannett-2003a,Burgin-Klinger-2004}. It is
precisely due to the existence of Turing's oracle machines  that J. Copeland and D. Proudfoot introduced the term
`hypercomputation' by 1999~\cite{Copeland-Proudfoot-1999a} for to
replace the wrong expressions such as  `super-Turing computation',
`computing beyond Turing's limit', and `breaking the Turing barrier',
and similar.

Recently Tien~D.~Kieu has proposed  an quantum algorithm to solve
the TM incomputable
\footnote{We follow to  S. B. Cooper y P. Odifreddi and we adopt the terminology  Turing's `computable' at replace of  Kleene `recursive' (see footnote 1 in~\cite{Cooper-Odifreddi-2003}).} 
problem named Hilbert's tenth problem, using as physical referent the well known simple harmonic
oscillator (SHO), which by effect of the second quantization has
as associated dynamical algebra the Weyl-Heisenberg algebra
denoted $\mathfrak{g}_{\mathrm{W\!-\!H}}$~\cite{Kieu-2003a, Kieu-2003b, Kieu-2003c, Kieu-2004a, Kieu-2005,
Kieu-2005a}. From the  algebraic analysis of Kieu's
hypercomputational quantum algorithm (KHQA), we have identified
the underlying properties of the $\mathfrak{g}_{\mathrm{W\!-\!H}}$ algebra which are necessary (but not
sufficient) to  guaranty  KHQA works. Such properties are that
the dynamical algebra admits infinite-dimensional
irreducible representations with naturally associated coherent states.

The importance of KHQA inside the field of hypercomputation,
at the same tenor that the importance of hypercomputation within
the domain of computer science, can not be sub-estimated. This algorithm is a plausible
candidate for a practical implementation of the hypercomputation,
maybe within the scope of the quantum optics. The adaptation of
KHQA to a new physical referents different to the harmonic
oscillator, opens the possibility of analyze news viable
alternatives for its practical implementation more beyond of
quantum optics, maybe using quasi-particles of condensed matter
systems.

In this work we present an algebraic adaptation of KHQA, it is
to say, we present an hypercomputation model \emph{à la} Kieu, based on the selection of a dynamical algebra
which is different to the $\mathfrak{g}_{\mathrm{W\!-\!H}}$ algebra. We have selected the Lie algebra
$\mathfrak{su}(1,1)$, due to that this algebra posses the necessary
characteristics for to realize the hypercomputation and also due
to that such algebra has been identified as the dynamical algebra
associated to many relatively simple quantum systems.

More in concrete, the $\mathfrak{su}(1,1)$ algebra posses four kinds of
infinite-dimensional unitary irreducible representations (UIR): the positive discrete series, the negative discrete series,
the principal series and the complementary series~\cite{Groenevelt-Koelink-2002}. In this work we use only the
positive discrete series. From the other side, the $\mathfrak{su}(1,1)$ algebra admits different
kinds of coherent states such as Barut-Girardello, Perelomov,
nonlinear, and minimum uncertain~\cite{Wang-2000,Fu-Sasaki-1996}.
Beside of all these, the $\mathfrak{su}(1,1)$ algebra admits different
kinds of realizations. Within the field of quantum
optics we have realizations on systems with one, two and four
photon modes~\cite{Wang-2000,Fu-Sasaki-1996}, or with systems
such as the density-dependent Holstein-Primakoff~\cite{Wang-2000}. Within the domain of condensed matter, we have realizations on the following quantum potentials: infinite square well, the Pöschl-Teller potentials~\cite{Antoine-Gazeaub-Monceauc-Klauder-Penson-2001} and Calogero-Sutherland model~\cite{Fu-Sasaki-1996}. Other realizations from the $\mathfrak{su}(1,1)$ algebra
  arises from the mathematical physics at relation with the
  recursive properties of the special functions, namely Laguerre
oscillators~\cite{Borzov-2000,Borzov-Damaskinsky-2002,Jellal-2002}, Legendre
and Chebyshev oscillators~\cite{Borzov-Damaskinsky-2002}, Meixner
Oscillators~\cite{Groenevelt-Koelink-2002,Atakishiyev-Jafarov-Nagiyev-Wolf-1998}, and so on.

The present paper is realized at the following way. In the section
2 we introduce KHQA in such way that the algebraic issues have
been empathized and we explicit the hypercomputational
characteristics of the $\mathfrak{g}_{\mathrm{W\!-\!H}}$ algebra. In the section 3 based on the
analysis of such algebraic characteristics we show the general
structure and the mathematical properties of our  adaptation of
KHQA using the $\mathfrak{su}(1,1)$ algebra. In the section 4 we
note that the infinite cylindrical well and a modified
cylindrical well also admit a  realization of the $\mathfrak{su}(1,1)$
algebra. Moreover, based on the adaptation of KHQA
that we have realized using the infinite square well~\cite{Sicard-Velez-Ospina-2005,Sicard-Ospina-Velez-2005}, we
show new adaptations of KHQA for some of the physical referents
previously listed. Finally we present some conclusions.

\section{Kieu's hypercomputational quantum algorithm}\label{sec-Kieu-algorithm}
With base on the SHO and its associated dynamical algebra
$\mathfrak{g}_{\mathrm{W\!-\!H}}$, Kieu has
proposed an possible algorithm for the solution of the Hilbert's
tenth problem by the use of three  strategies: (i) Codification of
the instance of the Hilbert's tenth problem to solve, (ii) The
utilization of a non-standard version of quantum computation, and
(iii) The establishment of a halting criterion. The strategy (i)
has a background the occupation-number operator associated to the
$\mathfrak{g}_{\mathrm{W\!-\!H}}$ algebra. The
strategy (ii) is based on the adiabatic quantum
computation~\cite{Farhi-Goldstone-Gutman-Spiser-2000,Farhi-Goldstone-Gutman-Lapan-Lundgren-Preda-2001}
applied to  unbounded Hamiltonians, it is to say, this strategy
constitutes an application of the quantum adiabatic theorem for
the case of unbounded
operators~\cite{Messiah-1990,Avron-Elgart-1999}. The adiabatic
initialization is obtained with the aid of the coherent states and
the ladder operators which are associated to the dynamical algebra
$\mathfrak{g}_{\mathrm{W\!-\!H}}$. The strategy
(iii) demands a property to the initial state of the
adiabatic evolution. Such property is based on the probability
distribution associated to the coherent states corresponding to
the $\mathfrak{g}_{\mathrm{W\!-\!H}}$ algebra.
We now present at detail, every one of the strategies previously
enunciated, at such way that the possible algebraic
generalizations can arise easily.

\subsection{Mathematical background}
The  mathematical background  underlying to KHQA is
shown by the equations \eref{eq-10} and corresponds to the
mathematical formalism of the SHO within the formulation of the
second quantization. At \eref{K0} we introduce the Fock
occupation-number states denoted $\mathfrak{F}^{\mathrm{SHO}}$, where $\mathbb{N} = \{0, 1, 2, \ldots \}$ is the set of non-negative integers. At \eref{K1} the annihilation and creation operators $a$ and
$a^{\dagger}$ are introduced. The commutation
relations between the ladder operators are presented in \eref{K2}.
At \eref{K4} the spectral equation for the SHO is shown in
terms of the  Hamiltonian  $H^{\mathrm{SHO}}$ and of the energy levels
$E_{n}^{\mathrm{SHO}}$. At \eref{K5} the Hamiltonian $H^{\mathrm{SHO}}$
is given in terms of the ladder operators. At \eref{K6} is
presented the definition of the occupation-number operator $N^{\mathrm{SHO}}$
whose eigenvalues are denoted $n$ and  which will be crucial for
that follows. \Eref{K7} gives the definition and
the explicit form of the coherent states denoted
$\ket{\alpha}^{\mathrm{SHO}}$. Finally \eref{K8} shows
the Poisson form of the probability density for the random
variable $n$ corresponding to the coherent states \eref{K7}.

\begin{subequations}\label{eq-10}
\begin{equation}\label{K0}
    \mathfrak{F}^{\mathrm{SHO}} = \{\ket{n} \mid n \in \mathbb{N}\},
\end{equation}
\begin{equation}\label{K1}
    a \ket{0}=0, \qquad a \ket{n}=\sqrt{n}\ket{n-1}, \qquad  a^{\dagger} \ket{n}=\sqrt{n+1}\ket{n+1},
\end{equation}
\begin{equation}\label{K2}
    [a,a^{\dagger}] = \oneI
\end{equation}
\begin{equation}\label{K4}
 H^{\mathrm{SHO}} \ket{n} = E_{n}^{\mathrm{SHO}}\ket{n},
\end{equation}
\begin{equation}\label{K5}
 H^{\mathrm{SHO}} = \hbar \omega^{\mathrm{SHO}}(a^{\dagger}a + 1/2),
\end{equation}
\begin{equation}\label{K6}
  N^{\mathrm{SHO}}= a^{\dagger}a, \quad N^{\mathrm{SHO}}\ket{n}  =n\ket{n},
\end{equation}
\begin{eqnarray}
    a\ket{\alpha}^{\mathrm{SHO}} &= \alpha\ket{\alpha}^{\mathrm{SHO}} \nonumber
    \\
    &= e^{-\frac{|\alpha|^{2}}{2}}\sum_{n=0}^{\infty}\frac{\alpha^{n}}{\sqrt{n!}}\ket{n}, \mbox{ where }  \alpha \in \mb{C}, \label{K7}
\end{eqnarray}
\begin{equation}\label{K8}
    P_{n}^{\mathrm{SHO}}(\alpha) =  e^{-|\alpha|^{2}}\frac{|\alpha|^{2n}}{n!}.
\end{equation}
\end{subequations}

\subsection{First strategy: Hilbert's tenth problem and its codification}
A Diophantine equation is an equation of the form
\begin{equation}\label{eq-20}
D(x_1, \ldots, x_k) = 0,
\end{equation}
where $D$ is a polynomial with integer coefficients. By 1990,
David Hilbert presented  his  famous  list of 23 problems. From
such list we extract the problem  number 10. In present
terminology, Hilbert's tenth problem may be paraphrased as:
\begin{quote}
  Given a Diophantine equation with any number of unknowns:
  To devise a process according to which it can be determined by a finite number of operations whether the equation has non-negative integers solutions.
\end{quote}

From the concluding results obtained by Matiyasevich, Davis,
Robinson, and Putnam, we know actually  that, at the general case,
this problem is algorithmically insolvable or more precisely, it
is TM incomputable~\cite{Matiyasevich-1993}. Justly, the possible
hypercomputability of Kieu's algorithm is due to the fact that
this algorithm solves Hilbert's tenth problem.

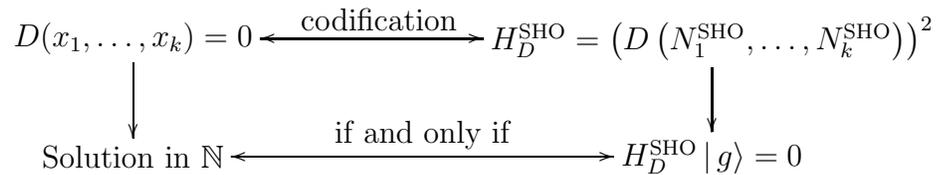
\begin{figure}[htbp]
\centering
\[\xymatrix{
        D(x_1, \ldots, x_k) = 0 \ar[rrr]^(0.425){\txt{codification}} \ar[d] &&& H_D^{\mathrm{SHO}} = \left(D \left ( N_1^{\mathrm{SHO}}, \ldots, N_k^{\mathrm{SHO}} \right )\right)^2 \ar[lll] \ar[d]
        \\
        \txt{Solution in $\mb{N}$} \ar[rrr]^{\txt{if and only if}} &&& H_D^{\mathrm{SHO}}\ket{g} = 0 \ar[lll]
      }\]
\vspace*{13pt}
\caption{\label{fig-10}Kieu's codification.}
\end{figure}

With the mathematical background that was presented at
\eref{eq-10}, Kieu proposes the codification of the  Hilbert's
tenth problem which is presented at the figure \ref{fig-10}. Such
figure illustrate that a  Diophantine equation of the kind
\eref{eq-20} is codified on a Hamiltonian denoted
$H_{\mathrm{D}}$ which results from the substitution of every unknown
in \eref{eq-20} by the  occupation-number operator defined in
\eref{K6}. At such way, the problem to determine if \eref{eq-20}
has solutions within the non-negative integers, is equivalent to
the problem to determine if the  energy associated to the
fundamental state denoted  $\ket{g}$, of the Hamiltonian
$H_{\mathrm{D}}$ is  zero.

\subsection{Second strategy: quantum adiabatic computation}
Due to the codification showed by the figure \ref{fig-10}, is
necessary to use a strategy of quantum computation which is
different to the  standard quantum computation (based on sequences
of unitary quantum logic gates that process
qubits)~\cite{Chuang-Nielsen-2000}. At words of Kieu we present
the strategy of quantum computation in the following form~\cite[p.
7]{Kieu-2005}:

\emph{In general, it is much more difficult to construct a specific state
for a quantum mechanical system than to control the physical process
(that is, to create a corresponding Hamiltonian) to which the system
is subject. One systematic method to obtain the ground state of a
Hamiltonian is to exploit the quantum adiabatic theorem to reach the
desired state through some adiabatic evolution which starts from a
readily constructible ground state of some other Hamiltonian. This
is the idea of quantum adiabatic computation (QAC)~\cite{Farhi-Goldstone-Gutman-Spiser-2000} \dots}

\emph{In QAC, we encode the solution of our problem to the ground state of
some specific Hamiltonian. As it is easier to implement controlled
dynamical processes than to obtain the ground state, we start the
computation with the system prepared in a different but readily
obtainable ground state of some other Hamiltonian. This initial
Hamiltonian is then slowly extrapolated into the Hamiltonian whose
ground state is the desired one. The adiabatic theorem of quantum
mechanics (QAT)~\cite{Messiah-1990} stipulates that if the extrapolation
rate is sufficiently slow compared to some intrinsic scale, the
initial state will evolve into the desired ground state with a high
probability \dots Measurements then take place
finally on the system in order to identify the ground state, from
which the solution to our problem emerges \dots}

\emph{Now, to carry out a QAC for a given Diophantine
  equation \eref{eq-20}, we prepare our quantum mechanical system
in the readily constructible initial ground state
\begin{equation}\label{eq-30}
   \ket{g_I}^{\mathrm{SHO}} = \bigotimes_{i=1}^{k}\ket{\alpha_{i}}^{\mathrm{SHO}},
\end{equation}
of a {\em universal} (that is, independent of the given Diophantine
equation) initial Hamiltonian $H_I$, with some complex numbers
$\alpha$'s,
\begin{equation}\label{eq-40}
    H_I^{\mathrm{SHO}}=\sum_{i=1}^{k}(a^{\dagger}_{i}-\alpha_{i}^{*})(a_{i}-\alpha_{i}).
\end{equation}
This is just the Hamiltonian for shifted simple harmonic oscillators
whose ground state is the well-known coherent state in quantum
optics. We then subject the system to the time-dependent Hamiltonian
$H_{A}^{\mathrm{SHO}}$, which linearly extrapolates the initial Hamiltonian $H_I$
to the final Hamiltonian $H_D$ in a time interval $T$,
\begin{equation}\label{eq-50}
    H_A^{\mathrm{SHO}}(t)=\left ( 1-\frac{t}{T} \right ) H_{I}^{\mathrm{SHO}}+\frac{t}{T}H_{D}^{\mathrm{SHO}}.
\end{equation}
}

\subsection{The algorithm}
With base on the two mentioned strategies and the strategy that
will be presented at the following section, given a Diophantine
equation with $k$ unknowns of type \eref{eq-20}, Kieu provides
the following quantum algorithm to decide whether this equation
has any non-negative integer solution or not~\cite{Kieu-2003c,Kieu-2005}:

\begin{ssfenumerate}
\item Construct a physical process in which a system initially
starts with a direct product of $k$ coherent states
\begin{equation*}
\ket{\psi(0)} = \ket{g_I}^{\mathrm{SHO}},
\end{equation*}
and in which the system is subject to
a time-dependent Hamiltonian $H_A^{\mathrm{SHO}}(t)$ of \eref{eq-50} over the time
interval $[0,T]$, for some time $T$.

\item Measure through the time-dependent Schr{\"o}dinger equation
\begin{equation*}
  i\partial_t \ket{\psi(t)} = H_A^{\mathrm{SHO}}(t) \ket{\psi(t)}, \mbox{ for $t \in [0,T]$}
\end{equation*}
the maximum probability to find the system in a
particular occupation-number state at the chosen time $T$,
\begin{eqnarray*}
P(T)  &= \max_{\ket{\{n\}}} \left | \braket{\psi(T)}{\{n\}}  \right |^2
\\
&= \left | \braket{\psi(T)}{\{n\}^0} \right | ^2 ,
\end{eqnarray*}
where $\ket{\{n\}} = \bigotimes_{i=1}^k\ket{n_i}$, and
$\ket{\{n\}^0}$ is the maximum-probability number state with a
particular $k$-tuple $(n^0_1, \ldots, n^0_k)$.

\item If $P(T) \le 1/2$, increase $T$ and repeat all the steps above.
\item If
  \begin{equation}\label{eq-60}
    P(T) > 1/2
  \end{equation}
then $\ket{\{n\}^0}$ is the ground state of $H_D^{\mathrm{SHO}}$
(assuming
no degeneracy) and we can terminate the
algorithm and deduce a conclusion from the fact that $H_D^{\mathrm{SHO}}\ket{\{n\}^0} = 0$ iff equation \eref{eq-20} has a non-negative integer solution.
\end{ssfenumerate}

\subsection{Third strategy: the halting criterion}
One of the most common miss-understandings about to KHQA is
linked with the halting criterion of the algorithm. Some authors
claim that the QAT only establish the existence of a time of
execution of the algorithm which is finite but unknown and for
then there is not a verifiable halting criterion of the algorithm.
From the very early versions of the  algorithm, Kieu has been
alert of this situation and he has proposed the following halting
criterion~\cite[p. 9-11]{Kieu-2005}:

\emph{ Nevertheless, it is important to note that QAT is {\em not
constructive}, as with most theorems involving limiting processes.
It only tells us that for ``sufficiently large" $T$ the system is
``mostly" in the instantaneous eigenstate. But the theorem tells
us nothing quantitatively about the degrees of being
``sufficiently large" or ``mostly" \dots }

\emph{
In other words, QAT can only guarantee that the ground state is
achievable in a finite time interval but cannot specify what that
interval should be. That is, it cannot by itself give us any
indication when the ground state has been obtained so that the
algorithm can then be terminated at that point. For that, we need
another criterion \dots
}

\emph{
The crucial step of any quantum adiabatic algorithm is the
identification of the ground state of the final Hamiltonian, $H_D$.
In our case we do not in advance know in general how long is
sufficiently long (the Theorem offers no direct help here); all we
can confidently know is that for each Diophantine equation and each
set of $\alpha_i$'s there is a {\em finite} evolution time after
which the adiabaticity condition is satisfied. We thus have to find
another criterion to identify the ground state.
}

\emph{
The identification criterion we have found can be stated as:
The ground state of $H_D^{\mathrm{SHO}}$ is the Fock state $\ket{\{n\}^0}$
measuredly obtained with a probability of more than ${1}/{2}$ after
the evolution for some time $T$ of the initial ground state
$\ket{g_I}^{\mathrm{SHO}}$ according to the Hamiltonian \eref{eq-50}:
\begin{center}
$\ket{\{n\}^0}$ is the ground state of $H_D^{\mathrm{SHO}}$ if
$\left| \braket{\psi(T)}{\{n\}^0} \right |^2 > {1}/{2}$, for some
$T$,
\end{center}
provided the initial ground state $\ket{g_I}^{\mathrm{SHO}}$ of $H_I^{\mathrm{SHO}}$ does not
have any dominant component in the occupation-number eigenstates
$\ket{\{n\}}$ of $H_D^{\mathrm{SHO}}$,
\begin{equation}\label{eq-70}
\left | ^{\mathrm{SHO}}\braket{g_I}{\{n\}} \right | ^2 \le {1}/{2},
\forall\, \{n\};\end{equation} and provided that for $0 < t < T$,
\begin{equation}\label{eq-80}
  \braxket{e(t)}{H_D^{\mathrm{SHO}} - H_I^{\mathrm{SHO}}}{g(t)} \neq 0,
\end{equation}
where $\ket{g(t)}$ and $\ket{e(t)}$ are, respectively, the
instantaneous ground state and the first excited state of
$H_A^{\mathrm{SHO}}$ at the time $t$.}\footnote{The criterion \eref{eq-80} was added recently by Kieu~\cite{Kieu-2005,Kieu-2005b} to correct the finite-dimensional counterexamples pointed out by Smith~\cite{Smith-2005}. On other hand, there was an open problem in relation with infinite-dimensional case. In personal communication, Kieu told us have found a mathematical proof that halting criterion \eref{eq-60} is a good identification for the ground state in this case.}

\subsection{Crucial properties of the $\mathfrak{g}_{\mathrm{W\!-\!H}}$ algebra for KHQA}
According to \eref{eq-10}, the dynamical algebra associated to
the SHO is the Lie algebra denoted $\mathfrak{g}_{\mathrm{W\!-\!H}}$, whose generators are the operators $a,
a^{\dagger}$ and $\oneI$. The $\mathfrak{g}_{\mathrm{W\!-\!H}}$ algebra admits a infinite-dimensional UIR which is established by the action of its
generators over the space $\mathfrak{F}^{\mathrm{SHO}}$ and which is given
by \eref{K1}. From this representation the occupation-number
operator $N^{\mathrm{SHO}}$ is obtained whose spectrum coincides with
the non-negative integers $\mb{N}$ as is showed by \eref{K6}.
This spectrum is precisely the searching space of the solution,
associated to every one of the variables of \eref{eq-20}  and
justifies the strategy of codification which is showed at the
figure  \ref{fig-10}.

From the other side, the adiabatic initialization for KHQA
which is represented by \eref{eq-30} and  \eref{eq-40} comes
from the $\mathfrak{g}_{\mathrm{W\!-\!H}}$
algebra. The initial state $\ket{g_{I}}^{\mathrm{SHO}}$  is the direct
product of $k$ coherent states of the  form \eref{K7}, and the
initial Hamiltonian denoted $H_{I}^{\mathrm{SHO}}$ is constructed
starting from the ladder operators $a^{\dagger}$ and $a$ of the
$\mathfrak{g}_{\mathrm{W\!-\!H}}$ algebra.
Besides of this, the identification of the ground state of
$H_D^{\mathrm{SHO}}$  that assumes the role of halting criterion for
the algorithm  according to \eref{eq-60} is supported on the
condition \eref{eq-70} which is satisfied by the probability
density $P_{n}^{\mathrm{SHO}}$ of  \eref{K8}. In concrete, the chosen of the coherent state with the form
\eref{eq-30} as the initial ground state
  entails the condition \eref{eq-70}, since for any $\alpha \neq 0$, and $\forall n > 0$
\begin{equation*}
  \left | \braket{\alpha}{n} \right |^2 = P_n^{\mathrm{SHO}}(\alpha) < 1/2.
\end{equation*}

\section{Hypercomputational quantum algorithm based on the algebra $\mathfrak{su}(1,1)$}\label{sec-adaptation}
From the algebraic point of view, the peculiarities of the algebra
$\mathfrak{g}_{\mathrm{W\!-\!H}}$ which are
required by KHQA,  start from the fact that this algebra
admits a infinite-dimensional  UIR that  operates over
the Fock space and its corresponding coherent states. With base on
such infinite-dimensional UIR is possible to establish
the needed ladder operators that let the construction of a number
operator and the associated coherent states, being all that the
basic algebraic ingredients of KHQA.

Due to the fact that the $\mathfrak{g}_{\mathrm{W\!-\!H}}$ algebra is not the only dynamical algebra that
satisfies the needed algebraic conditions, arises then the problem
of the adaptation of KHQA to other dynamical algebras and for
then to other physical systems.

We present in this section the adaptation of KHQA to the case
of the $\mathfrak{su}(1,1)$ algebra. Such algebra is chosen due to the
fact that this algebra is the dynamical algebra associated with
many well known physical systems.

The algebra $\mathfrak{su}(1,1)$ is defined by the commutation relations
\begin{equation}\label{eq-90}
   [K_{0},K_{1}]=  iK_{2}, \qquad [K_{0},K_{2}] = -K_{1}, \qquad [K_{1},K_{2}] = -iK_{0},
\end{equation}
or by  the commutation relations
\begin{equation}\label{eq-100}
   [K_{0},K_{\pm}]=  \pm K_{\pm}, \qquad [K_{+},K_{-}] = -2K_{0}, \mbox{ where } K_{\pm} \equiv (K_1 \pm iK_2).
\end{equation}

In contrast with the $\mathfrak{g}_{\mathrm{W\!-\!H}}$ algebra, the algebra $\mathfrak{su}(1,1)$ admits
different kinds of coherent states besides of various kinds of
representations. Here we use the named positive discrete
representation, which is defined as~\cite{Fu-Sasaki-1996,Wang-2000}
\begin{eqnarray*}
    K_{-}\ket{k,n} &= \sqrt{n(2k + n -1)}\ket{k,n-1},
    \\
    K_{+}\ket{k,n} &= \sqrt{(n+1)(2k +n)}\ket{k,n+1},
    \\
    K_{3}\ket{k,n} &=  (n+k)\ket{k,n},
\end{eqnarray*}
where $\ket{k,n} (n \in \mb{N})$ is the normalized  basis and $k
\in \left \{ \frac{1}{2}, 1, \frac{3}{2}, 2, \dots \right \}$ is
the Bargmann index labeling the UIR\footnote{Following to
 Antoine \emph{et al.} we recall that $k \in \left \{
\frac{1}{2}, 1, \frac{3}{2}, 2, \dots  \right \}$ for the discrete
series \emph{stricto senso} UIRs of $\mathfrak{su}(1,1)$, and $k \in
[1/2,+\infty)$ for the extension to the universal covering of the
group $SU(1,1)$~\cite{Antoine-Gazeaub-Monceauc-Klauder-Penson-2001}. Whatever this the fact, we will speak about discrete series UIRs for both cases.}.

We introduce the number operator $N^{\mathfrak{su}(1,1)}$ by
\begin{equation}\label{eq-120}
      N^{\mathfrak{su}(1,1)} = K_{0} -k, \qquad  N^{\mathfrak{su}(1,1)}\ket{k,n} = n \ket{k,n},
\end{equation}
and a well know coherent states of $\mathfrak{su}(1,1)$, are the
denominated  Barut-Girardello coherent states (BGCS). The BGCS are
defined as the eigenstates of the lowering operator $K_{-}$
\begin{equation}\label{eq-130}
K_{-}\ket{k,\alpha} _{BG}=\alpha \ket{k,\alpha}_{BG},
\end{equation}
and theses can be expressed as~\cite{Wang-2000}
\begin{equation}\label{eq-140}
\ket{k,\alpha}_{BG} = \sqrt{\frac{|\alpha
|^{2k-1}}{I_{2k-1}(2|\alpha |)}} \sum_{n=0}^\infty \frac{\alpha
^n}{\sqrt{n!\Gamma (n+2k)}}\ket{k,n},
\end{equation}
where $I_\nu (x)$ is the first kind modified Bessel function.

All these is well known. Now, we introduce a generalizations
of equations ~\eref{eq-90} to \eref{eq-140} which we will use for
that follow. Assuming the existence of a  quantum system
$\mathrm{S}$, whose dynamical algebra is $\mathfrak{su}(1,1)$, the
Eqs.~\eref{eq-150} gives the adaptation of the equations
\eref{eq-10} for the case of $\mathfrak{su}(1,1)$. \Eref{CG0}
defines the the Fock space of quantum states which is denoted
$\mathfrak{F}^{\mathrm{S}}$ corresponding to a quantum system denoted
$\mathrm{S}$. \Eref{CG1} presents the commutation relations
that defines the $\mathfrak{su}(1,1)$ algebra, where the generators
$K_{+}^{\mathrm{S}}$ and $K_{-}^{\mathrm{S}}$ correspond respectively to
the creation and destruction operators of $\mathfrak{su}(1,1)$. \Eref{CG2} shown the action of the infinite-dimensional UIR
of $\mathfrak{su}(1,1)$ over the space $\mathfrak{F}^{\mathrm{S}}$, with
characteristic function $f^{\mathrm{S}}$ which is assumed as quadratic
function of $n$. The expression \eref{CG3} shows the equation of
the energy spectrum for the quantum system and \eref{CG4} gives
the factorized form of the Hamiltonian $H^{\mathrm{S}}$ in terms of
the ladder operators. \Eref{CG5} defines the number
operator associated to the system  $\mathrm{S}$ and its corresponding
action over the space $\mathfrak{F}^{\mathrm{S}}$. The equation \eref{CG8}
is a definition of a non-linear coherent state denoted  $\ket{z}$
which is a generalization of the more standard linear coherent
states of the Barut-Girardello and
Klauder-Perelomov~\cite{Wang-2000,Wang-2000a} kinds; and \eref{CG9} presents the form of such non-linear coherent
states (see Appendix A).  \Eref{CG10} shows the
probability density for the random variable $n$, which is
associated to the generalized  coherent states \eref{CG9}.

\begin{subequations}\label{eq-150}
\begin{equation}\label{CG0}
    \mathfrak{F}^{\mathrm{S}} =  \{\ket{n} \mid n \in \mathbb{N}\},
\end{equation}
\begin{equation}\label{CG1}
  [K_{-}^{\mathrm{S}},K_{+}^{\mathrm{S}}]=  K_{3}^{\mathrm{S}}, \qquad [K_{-}^{\mathrm{S}},K_{3}^{\mathrm{S}}] = 2K_{-}^{\mathrm{S}}, \qquad [K_{+}^{\mathrm{S}},K_{3}^{\mathrm{S}}] = -2K_{+}^{\mathrm{S}},
\end{equation}
\begin{equation}\label{CG2}
\eqalign{
  K_{-}^{\mathrm{S}} \ket{0} &= 0, 
  \cr  
  K_{-}^{\mathrm{S}} \ket{n} &= \sqrt{f^{\mathrm{S}}(n)}\ket{n-1},
  \cr
   K_{+}^{\mathrm{S}} \ket{n} &= \sqrt{f^{\mathrm{S}}(n+1)}\ket{n+1},
  \cr
    K_{3}^{\mathrm{S}}\ket{n} &= \left ( f^{\mathrm{S}}(n+1)-f^{\mathrm{S}}(n) \right )\ket{n} = g^{\mathrm{S}}(n)\ket{n},
}
\end{equation}
\begin{equation}\label{CG3}
    H^{\mathrm{S}} \ket{n} = E_{n}^{\mathrm{S}}\ket{n},
\end{equation}
\begin{equation}\label{CG4}
H^{\mathrm{S}} = \hbar \omega(K_{+}^{\mathrm{S}}K_{-}^{\mathrm{S}}), \mbox{ with } K_{+}^{\mathrm{S}}K_{-}^{\mathrm{S}}\ket{n}= f^{\mathrm{S}}(n)\ket{n},
\end{equation}
\begin{equation}\label{CG5}
\eqalign{
N^{\mathrm{S}} &= \left( f^{\mathrm{S}} \left ( H^{\mathrm{S}} \right ) \right )^{-1} = \left( g^{\mathrm{S}} \left ( K_{3}^{\mathrm{S}} \right ) \right )^{-1},
\cr
N^{\mathrm{S}}\ket{n} &= n \ket{n},
}
\end{equation}
\begin{equation}\label{CG8}
h^{\mathrm{S}}(N^{\mathrm{S}})K_{-}^{\mathrm{S}} \ket{z}^{\mathrm{S}} = z
\ket{z}^{\mathrm{S}}, z \in \mb{C},
\end{equation}
\begin{equation}\label{CG9}
\fl \ket{z}^{\mathrm{S}} = \left ( \sum_{m=0}^{\infty}
  \frac{|z|^{2m}}{\left ( \prod_{j=0}^{m-1}h^{\mathrm{S}}(j)\right )^{2} \left ( f^{\mathrm{S}}(m)! \right )} \right )^{-1/2} \sum_{n=0}^{\infty}\frac{z^{n}}{\left( \prod_{j=0}^{n-1}h^{\mathrm{S}}(j) \right ) \left ( \sqrt{f^{\mathrm{S}}(n)!} \right )},
\end{equation}
\begin{equation}\label{CG10}
\fl     P_{n}^{\mathrm{S}}(z) = \left (\sum_{m=0}^{\infty}
\frac{|z|^{2m}}{\left( \prod_{j=0}^{m-1}h^{\mathrm{S}}(j) \right )^{2} \left ( f^{\mathrm{S}}(m)! \right ) } \right )^{-1} \frac{|z|^{2n}}{\left( \prod_{j=0}^{n-1}h^{\mathrm{S}}(j) \right )^{2} \left ( f^{\mathrm{S}}(n)! \right )}
\end{equation}
\end{subequations}

Starting from the Eqs.~\eref{eq-150}, the annunciated
adaptation of KHQA for the case of the $\mathfrak{su}(1,1)$ algebra
is completely direct. In the strategy of codification which was
represented by the figure \ref{fig-10}, it is necessary to
replace the Hamiltonian $H_D^{\mathrm{SHO}}$ by a new  Hamiltonian
denoted $H_D^{\mathrm{S}}$ which is constructed  with the number
operators defined at \eref{CG5}, by means of
  \begin{equation}\label{eq-160}
H_D^{\mathrm{S}} = \left( D \left ( N_1^{\mathrm{S}}, \ldots, N_k^{\mathrm{S}} \right ) \right)^2.
  \end{equation}
The adiabatic initialization  is obtained from the coherent states
\eref{CG9} and from \eref{CG8} and it is given by
\begin{eqnarray}
   \ket{g_{I}}^{\mathrm{S}} &= \bigotimes_{i=1}^{k}\ket{z_{i}}^{\mathrm{S}},
    \label{eq-170}
\\
H_I^{\mathrm{S}} &= \sum_{i=1}^{k}\left (
K_{+_i}^{\mathrm{S}}h^{\mathrm{S}}(N^{\mathrm{S}}) -z_{i}^{*} \right ) \left
(h^{\mathrm{S}}(N^{\mathrm{S}})K_{-_i}^{\mathrm{S}} - z_i \right).
\label{eq-180}
\end{eqnarray}
From all this, we obtain the Hamiltonian denoted $H_A^{\mathrm{S}}$
which is the generator of the adiabatic evolution and which is of
the form
\begin{equation}\label{eq-190}
     H_A^{\mathrm{S}}(t)=\left ( 1-\frac{t}{T} \right ) H_I^{\mathrm{S}}+\frac{t}{T}H_D^{\mathrm{S}}.
\end{equation}

Finally, to satisfy the halting criterion \eref{eq-60}, we chose
a value for the parameter $z \in \mb{C}$ which according to \eref{CG10} satisfies the condition in  \eref{eq-70}, it is to
say we chose a value of $z$ such that
\begin{equation*}
 P_{n}^{\mathrm{S}}(z) < 1/2.
\end{equation*}

All that is the abstract generalization or extension of the
original hypercomputational quantum algorithm of Kieu. In
the following section we will present some concrete physical
referents on which to realize the implementation of the abstract
algorithm previously presented.

\section{ Adaptation of KHQA over some concrete physical referents}
In this section some concrete quantum systems are presented as
possible physical referents on which to try to implement the
adaptation of KHQA with the $\mathfrak{su}(1,1)$ algebra, which was
presented at the past section. The mathematical adaptation
expressed by the Eqs.~\eref{eq-150} depends on the particular
forms of the characteristic functions $f^{\mathrm{S}}$ and
$h^{\mathrm{S}}$ associated with the physical system $\mathrm{S}$. Then
for every one of the considered physical systems, we establish
that the corresponding dynamical algebra is precisely
$\mathfrak{su}(1,1)$ and we determine the particular forms of
$f^{\mathrm{S}}$ y $h^{\mathrm{S}}$. The physical systems that are
considered here are: the infinite square well, the Pöschl-Teller
potentials, the infinite cylindrical well, a perturbed
cylindrical well, the density-dependent Holstein-Primakoff system
of quantum optics and the Laguerre oscillator. Other
systems of quantum optics such as: two-mode realization,
amplitude-squared realization, four-mode system; admit also
infinite-dimensional representations but such representations are
reducible and with such kind of representations is more difficult to adapt KHQA..

\subsection{The infinite square well}
The adaptation of the KHQA for the case of  the infinite square
well (ISW) was realized by the present authors within a previous
work~\cite{Sicard-Velez-Ospina-2005,Sicard-Ospina-Velez-2005}. At
the present work  we again establish that the  ISW satisfy the
mathematical  structure given by \eref{eq-150}, for a
particular forms of $f^{\mathrm{ISW}}$ y $h^{\mathrm{ISW}}$ from which it
is possible to construct the constitutive elements of the basic
algebraic anatomy of the KHQA.

For a particle with mass $m$ which is trapped inside the infinite
square well $0 \leq x \leq \pi l$, the Fock space associated
$\mathfrak{F}^{\mathrm{ISW}}$, the Hamiltonian operator $H^{\mathrm{ISW}}$, the
eigenvalue equation and the boundary conditions
are~\cite{Antoine-Gazeaub-Monceauc-Klauder-Penson-2001}
\begin{eqnarray}
  \mathfrak{F}^{\mathrm{ISW}} &=  \{\ket{n} \mid n \in \mathbb{N}\}, \nonumber
\\
  H^{\mathrm{ISW}} &= i^2 \frac{\hbar^2}{2m}\frac{\rmd^2}{\rmd x^2} - \frac{\hbar^2}{2ml^2}, \label{eq-isw-10}
    \\
     H^{\mathrm{ISW}}\psi^{\mathrm{ISW}} &=  E^{\mathrm{ISW}}\psi^{\mathrm{ISW}}, \label{eq-isw-15}
\\
\psi(x) &=0, \quad x \geq \pi l \mbox{ and } x \leq 0. \label{eq-isw-17}
\end{eqnarray}

Replacing  \eref{eq-isw-10} on \eref{eq-isw-15} together with
the boundary conditions \eref{eq-isw-17}, we obtain
\begin{equation*}
  \psi_n^{\mathrm{ISW}}(x) = \sqrt{\frac{2}{\pi l}} \sin(x+1)\frac{x}{l} \equiv \braket{x}{n}, 
\end{equation*}
\begin{equation}\label{eq-isw-18}
\fl E_n^{\mathrm{ISW}} = \hbar \omega^{\mathrm{ISW}} e_n^{\mathrm{ISW}}, \mbox{ where } \omega^{\mathrm{ISW}} = \frac{\hbar^2}{2ml^{2}} \mbox{ and } e_n^{\mathrm{ISW}} = n(n+2), \quad n \in \mb{N}, 
\end{equation}
and where the action of $H^{\mathrm{ISW}}$ over the space
$\mathfrak{F}^{\mathrm{ISW}}$ is given
\begin{equation*}
H^{\mathrm{ISW}} \ket{n} = E_n^{\mathrm{ISW}} \ket{n}.
\end{equation*}

Due to the spectral structure of the ISW, the dynamical algebra
associated with it, is the Lie algebra
$\mathfrak{su}(1,1)$~\cite{Antoine-Gazeaub-Monceauc-Klauder-Penson-2001}
 whose generators denoted $K_+^{\mathrm{ISW}}, K_-^{\mathrm{ISW}}$ and $K_3^{\mathrm{ISW}}$
 satisfy
 the commutation relations of \eref{CG1}. With base on  \eref{eq-isw-18},
 the algebra $\mathfrak{su}(1,1)$ admits an infinite-dimensional UIR over the space $\mathfrak{F}^{\mathrm{ISW}}$, which is given by
\begin{eqnarray*}
    K_{-}^{\mathrm{ISW}} \ket{0} &= 0,
    \\
     K_{-}^{\mathrm{ISW}} \ket{n} &= \sqrt{e_n^{\mathrm{ISW}}}\ket{n} = \sqrt{n(n+2)}\ket{n-1},
    \\
    K_{+}^{\mathrm{ISW}} \ket{n} &= \sqrt{e_{n+1}^{\mathrm{ISW}}}\ket{n} = \sqrt{(n+1)(n+3)}\ket{n+1},
\\
K_{3}^{\mathrm{ISW}}\ket{n} &= \left (e_{n+1}^{\mathrm{ISW}} - e_n^{\mathrm{ISW}} \right ) \ket{n} = (2n+3)\ket{n}.
\end{eqnarray*}

With basis on this representation of the algebra $\mathfrak{su}(1,1)$,
the Hamiltonian \eref{eq-isw-10} is rewritten as
\begin{equation*}
H^{\mathrm{ISW}} = \hbar \omega(K_{+}^{\mathrm{ISW}}K_{-}^{\mathrm{ISW}}), \qquad  H^{\mathrm{ISW}} \ket{n} = E_n^{\mathrm{ISW}} \ket{n},
\end{equation*}
and a new number operator $N^{\mathrm{ISW}}$ is given by
\begin{equation*}
 N^{\mathrm{ISW}} = (1/2)\left ( K_3^{\mathrm{ISW}} - 3 \right),  \qquad N^{\mathrm{ISW}}\ket{n} = n \ket{n}.
\end{equation*}

Due to the associated  dynamical algebra, the BGCS $\ket{z}^{\mathrm{ISW}}, z \in \mb{C}$, for the ISW
are given by~\cite{Wang-Sanders-Pan-2000}
\begin{equation}\label{eq-isw-20}
\fl  K_{-}^{\mathrm{ISW}} \ket{z}^{\mathrm{ISW}} = z \ket{z}^{\mathrm{ISW}}, \mbox{ where } \ket{z}^{\mathrm{ISW}} = \frac{\modulo{z}}{\sqrt{I_2(2\modulo{z})}}\sum_{n=0}^{\infty}\frac{z^n}{\sqrt{n!(n+2)!}}\ket{n},
\end{equation}
where $I_v(x)$ is the modified Bessel function of the first kind.
The corresponding probability density for the random variable $n$
that results from \eref{eq-isw-20} es
\begin{equation*}
   P_{n}^{\mathrm{ISW}}(z) = \frac{|z|^{2}}{I_{2}(2|z|)}\frac{|z|^{2n}}{n !
    (2+n)!}.
\end{equation*}

We have established then that the ISW satisfy the algebraic
structure of \eref{eq-150} where the characteristic functions
are of the forms
\begin{eqnarray*}
  f^{\mathrm{ISW}}(n) &= e_n^{\mathrm{ISW}} = n(n+2),
 \\
 h^{\mathrm{ISW}}(N^{\mathrm{ISW}}) &= \oneI,
\end{eqnarray*}
and we can on consequence to rewrite in terms of the ISW, all the
elements of the KHQA, which are given by the Eqs.~\eref{eq-160} to \eref{eq-190}; and to obtain at such way, an
adaptation of the  KHQA for the ISW, on where the halting criterion
\eref{eq-60}  is satisfied according to the condition
\eref{eq-70}  with every one value of $z \in \mb{Z}$ that
verifies that
\begin{equation*}
    P_{0}^{\mathrm{ISW}}(z)= \frac{|z|^{2}}{2I_{2}(2|z|)} < 1/2,
\end{equation*}
given that $P_{n}^{\mathrm{ISW}}(z) \leq P_{0}^{\mathrm{ISW}}(z), \forall
n$.

\subsection{The infinite cylindrical well and the perturbed cylindrical well}
With the aim to adapt the  KHQA over the infinite cylindrical well
(ICW) or over a perturbed cylindrical well (PCW), initially
we establish that these physical referents have as dynamical
algebra justly the $\mathfrak{su}(1,1)$ algebra. We follow the work that
was realized by Antoine \emph{et
al.}~\cite{Antoine-Gazeaub-Monceauc-Klauder-Penson-2001}, at such
way that we obtain the particular forms of the functions
$f^{\mathrm{ICW/PCW}}$ and  $h^{\mathrm{ICW/PCW}}$ that are required.

At concrete, for a particle with mass $m$ which is  trapped inside
the infinite cylindrical well of radius $R$, the Fock space
associated denoted $\mathfrak{F}^{\mathrm{ICW}}$ and the corresponding
Hamiltonian operator $H^{\mathrm{ICW}}$ are given by:
\begin{eqnarray}
   \mathfrak{F}^{\mathrm{ISW}} &=  \{\ket{n} \mid n \in \mathbb{N}\}, \label{eq-icw-00A0}
\\
    H^{\mathrm{ICW}} &= -\frac{\hbar ^{2}}{2m}\nabla ^{2} + U^{\mathrm{ICW}}, \label{eq-icw-00A}
\end{eqnarray}
where $U^{\mathrm{ICW}}$ is a constant which will be obtained
posteriorly and the bi-dimensional Laplacian operator is written
on cylindrical coordinates. The spectral equation for Hamiltonian $H^{\mathrm{ICW}}$
is
\begin{equation}\label{eq-icw-00B}
    H^{\mathrm{ICW}} \Psi^{\mathrm{ICW}} = E^{\mathrm{ICW}}\Psi^{\mathrm{ICW}}.
\end{equation}
Now the substitution of \eref{eq-icw-00A} on \eref{eq-icw-00B} gives the
following partial differential equation whose solution determines
the spectrum of $H^{\mathrm{ICW}}$
\begin{equation}\label{eq-icw-00}
    -\frac{\hbar ^{2}}{2m}\nabla ^{2}\Psi^{\mathrm{ICW}} + U^{\mathrm{ICW}} \Psi^{\mathrm{ICW}} =E^{\mathrm{ICW}}  \Psi^{\mathrm{ICW}}.
\end{equation}
Using cylindrical coordinates and axial symmetry \eref{eq-icw-00} is
reduced to
\begin{equation}\label{eq-icw-10}
    \frac{\partial ^{2}}{\partial r}\Psi^{\mathrm{ICW}}(r)+\frac{1}{r}\frac{\partial}{\partial
    r}\Psi^{\mathrm{ICW}}(r)+\frac{2m}{\hbar ^{2}}(E^{\mathrm{ICW}}- U^{\mathrm{ICW}})\Psi^{\mathrm{ICW}}(r)=0.
\end{equation}
The condition of trapping for the particle within the interior of
the ICW, is introduced using the boundary condition
\begin{equation}\label{eq-icw-20}
    \Psi^{\mathrm{ICW}}(R)=0.
\end{equation}
The solution of \eref{eq-icw-10} with the condition of wave
function finite at $r=0$ is given by
\begin{equation}\label{eq-icw-30}
    \Psi^{\mathrm{ICW}}(r)=C J_{0}\left ( \sqrt{\frac{2m(E^{\mathrm{ICW}} - U^{\mathrm{ICW}})}{\hbar ^{2}}}r \right),
\end{equation}
where $C$ is a constant. Now using \eref{eq-icw-30} and the
boundary condition \eref{eq-icw-20}, the energy spectrum is
obtained as
\begin{equation}\label{eq-icw-40}
    E_n^{\mathrm{ICW}} = U^{\mathrm{ICW}} + \frac{\hbar ^{2}}{2 m R^{2}}\alpha _{n}^{2},\end{equation}
where $n \in \mb{N}$ and the $\alpha_{n}$'s are the zeros of the Bessel function $J_{0}(x)$. Using the empirical formula of interpolation for $\alpha_{n}$
\begin{equation}\label{eq-icw-50}
   \alpha _{n}=3.115\,n+ 2.405,
\end{equation}
the substitution of \eref{eq-icw-50} on \eref{eq-icw-40} gives
\begin{eqnarray}
  E_{n}^{\mathrm{ICW}} &= U^{\mathrm{ICW}}+\frac{2.89 \hbar ^{2}}{m R^{2}}+\frac{4.85 \hbar ^{2}}{m R^{2}}n(n+1.54) \nonumber
   \\
&= \hbar \omega^{\mathrm{ICW}}e_n^{\mathrm{ICW}}, \label{eq-icw-60}
\end{eqnarray}
where $U^{\mathrm{ICW}}= -2.89 \hbar ^{2} /m R^{2}, \omega^{\mathrm{PCW}} = 4.85 \hbar /mR^{2}$, and  $e_n^{\mathrm{ICW}} = n(n+1.54), \quad n \in \mb{N}$. 

The normalized wave function is given by 
\begin{equation*}
    \Psi_{n}^{\mathrm{ICW}}(r)=\frac{1}{R
    \sqrt{\pi}}\frac{J_{0}(\frac{\alpha_{n}}{R}r)}{J_{1}(\alpha_{n})}\equiv
    \braket{r}{n},
\end{equation*}
and the action of $H^{\mathrm{ICW}}$  over the space
$\mathfrak{F}^{\mathrm{ISW}}$ being
\begin{equation*}
H^{\mathrm{ICW}} \ket{n} =\hbar \omega^{\mathrm{ICW}} e_n^{\mathrm{ICW}}\ket{n}.
\end{equation*}

From the other side, for the case of the PCW, we consider a
quantum particle that it is confined to the interior of a infinite
long cylinder of finite radius $R$ but now the interior of the
cylinder has a potential  of the form
\begin{equation}\label{eq-pcw-000}
    V{^\mathrm{PCW}}(r)=W{^\mathrm{PCW}} + \frac{U{^\mathrm{PCW}}}{r^{2}},
\end{equation}
where $U{^\mathrm{PCW}}$ and $W{^\mathrm{PCW}}$ are constants that we can
to determine ulteriorly and we assume that both the wall of
cylinder and the axis of the cylinder always obstruct  that the
particle resides on them, because both the wall and the axis are
maintained to infinite potential.

For a particle with mass $m$ trapped inside the PCW of radius $R$,
the Fock space associated denoted  $\mathfrak{F}^{\mathrm{PCW}}$, the
Hamiltonian operator $H^{\mathrm{PCW}}$, and the spectral equation are
given by
\begin{eqnarray}
  \mathfrak{F}^{\mathrm{PCW}} &=  \{\ket{n} \mid n \in \mathbb{N}\}, \label{eq-pcw-001}
\\
H^{\mathrm{PCW}} &= -\frac{\hbar ^{2}}{2m}\nabla ^{2} + V^{\mathrm{PCW}}, \label{eq-pcw-00A2}
\\
H^{\mathrm{PCW}} \Psi^{\mathrm{PCW}} &= E^{\mathrm{PCW}}\Psi^{\mathrm{PCW}},
\label{eq-pcw-00B2}
\end{eqnarray}
where again the bi-dimensional Laplacian operator is written on
cylindrical coordinates. The substitution of \eref{eq-pcw-00A2} with \eref{eq-pcw-000} on
\eref{eq-pcw-00B2} gives the following partial differential equation
whose solution determines the spectrum of $H^{\mathrm{PCW}}$
\begin{equation}\label{eq-pcw-002}
    -\frac{\hbar ^{2}}{2m}\nabla ^{2}\Psi^{\mathrm{PCW}} + V^{\mathrm{PCW}}\Psi^{\mathrm{PCW}} =E^{\mathrm{PCW}} \Psi^{\mathrm{PCW}}.
\end{equation}
Using again cylindrical coordinates and axial symmetry  \eref{eq-pcw-002} is reduced to
\begin{equation}\label{eq-pcw-102}
\fl    \frac{\partial ^{2}}{\partial r}\Psi^{\mathrm{PCW}}(r)+\frac{1}{r}\frac{\partial}{\partial
    r}\Psi^{\mathrm{PCW}}(r)+\frac{2m}{\hbar ^{2}} \left ( E^{\mathrm{PCW}}- W^{\mathrm{PCW}} -\frac{U^{\mathrm{PCW}}}{r^{2}} \right ) \Psi^{\mathrm{PCW}}(r)=0.
\end{equation}
The condition of trapping  for the particle to the interior of the
PCW but with a infinite potential  at $r=0$, is
\begin{equation*}
    \Psi^{\mathrm{PCW}}(R)=0, \quad \Psi^{\mathrm{PCW}}(0) = 0,
\end{equation*}
then the solution of \eref{eq-pcw-102} with the condition  of wave
function finite at $r=0$ is given by
\begin{equation}\label{eq-pcw-302}
    \Psi^{\mathrm{PCW}}(r)=C J_{\sqrt{\frac{2mU^{\mathrm{PCW}}}{\hbar ^{2}}}}\left( \sqrt{\frac{2m(E^{\mathrm{PCW}}-W^{\mathrm{PCW}})}{\hbar ^{2}}}r \right),
\end{equation}
where again $C$ is a constant.
Now to obtain in \eref{eq-pcw-302} the condition $ \Psi^{\mathrm{PCW}}(0)=0$, is necessary that
\begin{equation}\label{eq-pcw-362}
    \sqrt{\frac{2m U^{\mathrm{PCW}}}{\hbar ^{2}}}\geq 1, \mbox{ where we have chosen that } \sqrt{\frac{2m U^{\mathrm{PCW}}}{\hbar ^{2}}} = 1.
\end{equation}
With \eref{eq-pcw-362}, then \eref{eq-pcw-302} changes to
\begin{equation}\label{eq-pcw-372}
    \Psi^{\mathrm{PCW}}(r)=C J_{1}\left(\sqrt{\frac{2m(E^{\mathrm{PCW}}-W^{\mathrm{PCW}})}{\hbar ^{2}}}r\right).
\end{equation}
Using \eref{eq-pcw-372} and the boundary condition $\Psi^{\mathrm{PCW}}(R)=0$, the
energy spectrum is obtained as
\begin{equation}\label{eq-pcw-402}
    E_{n}^{\mathrm{PCW}}=W^{\mathrm{PCW}}+\frac{\hbar ^{2}}{2 m R^{2}}\alpha _{n}^{2},
\end{equation}
where $n \in \mb{N}$ and the $\alpha_{n}$'s are the zeros of the Bessel function $J_{1}(x)$.
Using the empirical formula of interpolation for $\alpha_{n}$
\begin{equation}\label{eq-pcw-502}
   \alpha _{n}=3.14n+ 3.83,
\end{equation}
the substitution of \eref{eq-pcw-502} on \eref{eq-pcw-402} gives
\begin{eqnarray}
    E_{n}^{\mathrm{PCW}} &= W^{\mathrm{PCW}} + \frac{7.34 \hbar ^{2}}{m R^{2}} + \frac{4.93 \hbar ^{2}}{m R^{2}} n(n+2.43) \nonumber
 \\
  &= \hbar \omega^{\mathrm{PCW}}e_n^{\mathrm{PCW}}, \label{eq-pcw-503}
\end{eqnarray}
where $W^{\mathrm{PCW}} = -7.34 \hbar ^{2}/m R^{2}, \omega^{\mathrm{PCW}} =4.93  \hbar / m R^{2}$, and $e_n^{\mathrm{PCW}} = n(n+2.43), \quad n \in \mb{N}$. 

Finally the normalized wave function is
\begin{equation*}
    \Psi_{n}^{\mathrm{PCW}}(r)=\frac{1}{R
    \sqrt{\pi}}\frac{J_{1}(\frac{\alpha_{n}}{R}r)}{J_{0}(\alpha_{n})}\equiv
    \braket{r}{n},
\end{equation*}
and the action of $H^{\mathrm{PCW}}$ over the space
$\mathfrak{F}^{\mathrm{PCW}}$ is given by
\begin{equation*}
H^{\mathrm{PCW}} \ket{n} =\hbar \omega^{\mathrm{PCW}} e_n^{\mathrm{ICW}}\ket{n}.
\end{equation*}

With the purpose of to avoid a very heavy notation, we define by
the rest of this subsection a new variable denoted $i$ that can to
take the values  $\mathrm{ICW}$ and $\mathrm{PCW}$, it is to say
\begin{equation*}
  i \in \{\mathrm{ICW}, \mathrm{PCW} \}.
\end{equation*}

With the aim to establish that the $\mathfrak{su}(1,1)$ algebra is the
dynamical algebra associated both to the ICW as to the PCW, we
follow the procedure that was presented in~\cite{Antoine-Gazeaub-Monceauc-Klauder-Penson-2001} for the case
of ISW, we introduce both a creation operator denoted $K_{+}^i$ as
a destruction operator denoted  $K_{-}^i$, at such way that we can
to rewrite the Hamiltonian $H^i$ as
\begin{equation*}
H^i = \hbar \omega^i\left ( K_{+}^iK_{-}^i \right ).
\end{equation*}
Besides of this, we introduce the operator
\begin{equation*}
K_{3}^i = \left [ K_{-}^i,K_{+}^i\right ],
\end{equation*}
in such form that the operators $K_{+}^i, K_{-}^i$ y $K_{3}^i$
satisfy the commutation relations \eref{CG1}.

With the aim to satisfy the requirements, with  base on \eref{eq-icw-60} and \eref{eq-pcw-503} we establish a
representation of the $\mathfrak{su}(1,1)$ algebra which is given by
\begin{eqnarray*}
    K_{-}^i \ket{0} &= 0,
    \\
     K_{-}^i \ket{n} &= \sqrt{e_n^i}\ket{n-1},
    \\
    K_{+}^i \ket{n} &= \sqrt{e_{n+1}^i}\ket{n+1},
    \\
    K_{3}^i\ket{n} &= \left (e_{n+1}^i - e_n^i \right ) \ket{n}.
\end{eqnarray*}

From  \eref{eq-icw-60} and \eref{eq-pcw-503} we
define
\begin{equation*}
  b^{\mathrm{ICW}} = 1.54 \mbox{ and } b^{\mathrm{ICW}} = 2.43,
\end{equation*}
and with  base on the  representation that was introduced we get a
number operator of the form
\begin{equation*}
  N^i = (1/2)\left ( K_3^i - (b^i+1) \right), \qquad N^i\ket{n} = n \ket{n}.
  \end{equation*}

The BGCS for the ICW and for the PCW
are the states denoted $\ket{z}^i$ that satisfy the equation
$K_{-}^i \ket{z}^i= z \ket{z}^i$ and which have the form
\begin{equation*}
    \ket{z}^i = \frac{|z|^{\left ( b^i \right )/2}}{\sqrt{I_{b^i}(2|z|)}} \sum_{n=0}^{\infty}\frac{z^{n}}{\sqrt{n !
    (b^i + n)!}}\ket{n},
\end{equation*}
being the associated probability density denoted $P_{n}^i(z)$
\begin{equation*}
       P_{n}^i(z) =\frac{|z|^{b^i }}{I_{b^i }(2|z|)}\frac{|z|^{2n}}{n !
    (b^i +n)!}
\end{equation*}

Then, for the systems ISW/PCW, we have established that they
satisfy the algebraic structure of \eref{eq-150} with
\begin{eqnarray*}
  f^i(n) &= e_n^i
 \\
 h^i(N^i) &= \oneI,
\end{eqnarray*}

and by the use of a procedure which is similar to the realized for
the ISW, we obtain an adaptation of the KHQA over the systems
ICW/PCW where the halting criterion  \eref{eq-60} is satisfied for
the values of $z \in \mb{C}$ such that
\begin{equation*}
     P_{0}^i = \frac{|z|^{b^i}}{(b^i)!I_{b^i}(2|z|)} < 1/2.
\end{equation*}

\subsection{The Pöschl-Teller potentials}
In this subsection is showed that the Pöschl-Teller Potentials
(PTP) also satisfies the algebraic structure given \eref{eq-150}
and for then it is possible to adapt the KHQA for the case of the
PTP. The problem to find both the energy spectrum as the wave
functions for a particle of mass $m$ which is confined within a
ISW, is generalized to the case when the particle is trapped by a
potential of the Pöschl-Teller kind~\cite{Antoine-Gazeaub-Monceauc-Klauder-Penson-2001}
\begin{equation*}
V_{\lambda, \kappa}^{PTP}(x)= \frac{1}{2} V_0^{PTP} \left(\frac{\lambda(\lambda-1)}{\cos^2 x/2l}+\frac{\kappa(\kappa-1)}{\sin^2 x/2l}\right),
\end{equation*}
where the parameters $\lambda,\kappa >1$, the coupling constant is
$V_0 > 0$ and the PTP is defined inside the domain $0 \leq x \leq
\pi l$.
The corresponding Hamiltonian is given by
\begin{equation}\label{eq-140PTP}
   H^{\mathrm{PTP}} = i^2 \frac{\hbar^2}{2m}\fracd{\rmd^2}{\rmd x^2}+\fracd{\hbar^2}{8ml^2}\left(\frac{\lambda(\lambda-1)}{\cos^2 x/2}+ \frac{\kappa(\kappa-1)}{\sin ^2x/2}\right) - \frac{\hbar^2}{8ml^2}(\lambda+\kappa)^2,
\end{equation}
where we chose by convenience $V_0=\hbar^{2}/4ml^{2}$~\cite{Antoine-Gazeaub-Monceauc-Klauder-Penson-2001}. The energy
spectrum and the corresponding eigenstates comes from the
solutions of the spectral equation
\begin{equation}\label{eq-141PTP}
H^{\mathrm{PTP}}\psi^{\mathrm{PTP}}(x)=E^{\mathrm{PTP}}\psi^{\mathrm{PTP}}(x),
\end{equation}
with the boundary conditions
\begin{equation}\label{eq-142PTP}
\psi^{\mathrm{PTP}}(0)=\psi^{\mathrm{PTP}}(\pi l)=0.
\end{equation}
Replacing  \eref{eq-140PTP} on \eref{eq-141PTP} together with the
boundary conditions \eref{eq-142PTP}, we obtain the normalized
wave functions and their corresponding eigenvalues~\cite{Antoine-Gazeaub-Monceauc-Klauder-Penson-2001}
\begin{eqnarray*}
\fl \Psi_n^{\mathrm{PTP}}(x)&= [c_n(\lambda,\kappa)]^{-1/2} \left( \cos{\frac{x}{2l}}\right )^{\lambda} \left( \sin{\frac{x}{2l}} \right )^{\kappa}  {_2F_1}\left(-n,n+\lambda+\kappa;\kappa+\frac{1}{2};\sin^2{\frac{x}{2a}} \right)
\\
\fl & \equiv \braket{x}{\eta,n},
\\
\\
E_n^{\mathrm{PTP}} &= \hbar\omega^{\mathrm{PTP}} e_n^{\mathrm{PTP}}(\lambda,\kappa),
\end{eqnarray*}
where  $[c_n(\lambda,\kappa)]^{-1/2}$ is normalization factor
which is given analytically when $\lambda$ and $\kappa$ are
positive integers, the function $_2F_1$  is a particular case of
the generalized hypergeometric function, and
\begin{equation}\label{eq-143PTP}
\omega^{\mathrm{PTP}} = \frac{\hbar}{2ml^{2}}, \quad e_n^{\mathrm{PTP}}(\lambda,\kappa) = n(n+ 2\eta -1), \quad \eta = \frac{\lambda + \kappa + 1}{2}.
\end{equation}
The action of the Hamiltonian $H^{\mathrm{PTP}}$ over the Fock space
defined as
\begin{equation*}
\mathfrak{F}^{\mathrm{PTP}} =  \{\ket{\eta,n} \mid n \in \mathbb{N} \},
\end{equation*}
is given by
\begin{equation*}
    H^{\mathrm{PTP}}\ket{\eta,n}=E_n^{\mathrm{PTP}}\ket{\eta,n}.
\end{equation*}

Due to the spectral structure of the PTP, its dynamical algebra is
again
$\mathfrak{su}(1,1)$~\cite{Antoine-Gazeaub-Monceauc-Klauder-Penson-2001},
whose generators denoted now,  $K_+^{\mathrm{PTP}}, K_-^{\mathrm{PTP}}$ and
$K_3^{\mathrm{PTP}}$ satisfy the commutation relations \eref{CG1}.
With base on \eref{eq-143PTP}, the $\mathfrak{su}(1,1)$ algebra admits
an infinite-dimensional UIR over the space
$\mathfrak{F}^{\mathrm{PTP}}$ which is given by
\begin{equation}\label{eq-144PTP}
 \eqalign{
    K_{-}^{\mathrm{PTP}} \ket{\eta,0} &= 0,
    \cr
     K_{-}^{\mathrm{PTP}} \ket{\eta,n} &= \sqrt{e_n^{\mathrm{PTP}}}\ket{\eta,n-1} = \sqrt{n(n+ 2\eta -1)}\ket{\eta,n-1},
    \cr
    K_{+}^{\mathrm{PTP}} \ket{\eta,n} &= \sqrt{e_{n+1}^{\mathrm{PTP}}}\ket{\eta,n+1} = \sqrt{(2\eta + n)(n+1)}\ket{\eta,n+1},
\cr
K_{3}^{\mathrm{PTP}}\ket{\eta,n} &= \left (e_{n+1}^{\mathrm{PTP}} - e_n^{\mathrm{PTP}} \right ) \ket{\eta,n} = (\eta + n)\ket{\eta,n}.
}
\end{equation}
With basis on  the representation of the $\mathfrak{su}(1,1)$ algebra,
the Hamiltonian $H^{\mathrm{PTP}}$ is rewritten as
\begin{equation*}
  H^{\mathrm{PTP}} = \hbar\omega^{\mathrm{PTP}} \left( K_+^{\mathrm{PTP}}K_-^{\mathrm{PTP}} \right), \end{equation*}
and we can to construct a new number operator of the form
\begin{equation*}
N^{\mathrm{PTP}} = (1/2)\left ( K_3^{\mathrm{PTP}} -\eta \right ), \quad  N^{\mathrm{PTP}}\ket{\eta,n} = n \ket{\eta,n}
\end{equation*}

The existence of the dynamical algebra permits the construction of
generalized coherent states to $\mathfrak{su}(1,1)$. The state
$\ket{z}, z \in \mb{C}$, is chosen again as of the
Barut-Girardello type; and again is defined as
$K_-^{\mathrm{PTP}}\ket{\eta,z}=z\ket{\eta,z}$. As is well know this is
a natural generalization of the coherent state associated to the
harmonic oscillator. The explicit form is
\begin{equation*}
\ket{\eta,z} =\left ( \Gamma(\eta)\modulo{z}^{-(\eta-1)}I_{\eta-1}(2\modulo{z})\right )^{-1/2} \sum_{n=0}^{\infty}\frac{z^n}{\sqrt{n!\,(\eta)_n}}\ket{\eta/2,n},
\end{equation*}
where $(\eta)_n$ is the Pochammer's symbol, defined as
$(\eta)_n=\eta(\eta+1)\cdots(\eta+n-1)$; and $I_{\eta-1}$ is again
the modified Bessel function of the first class. Similarly to the
previous systems, we have a probability density associated to the
coherent state which is immediately extracted from the explicit
form of the coherent state.

We have established then, that the PTP also satisfy the algebraic
structure given by \eref{eq-150} with characteristic functions
of the form
\begin{eqnarray*}
  f^{\mathrm{PTP}}(n) &= e_n^{\mathrm{PTP}}(\lambda,\kappa)
 \\
 h^{\mathrm{PTP}}(N^{\mathrm{PTP}}) &= \oneI,
\end{eqnarray*}
and by a procedure which is similar to the realized for the cases
of ISW, ICW and PCW, we obtain an adaptation of the KHQA for the
case of the  PTP where it is possible to chose the values of the
parameters in such way that halting criterion \eref{eq-60} is
satisfied.

\subsection{The density-dependent Holstein-Primakoff system}
In this subsection is showed that the density-dependent
Holstein-Primakoff (HP) system of quantum optics  also satisfies
the algebraic structure of \eref{eq-150}. The HP realization of
the Lie algebra $\mathfrak{su}(1,1)$ is constructed from the generators
denoted now, $K_{+}^{\mathrm{HP}}$, $K_{-}^{\mathrm{HP}}$ and
$K_{3}^{\mathrm{HP}}$; that satisfy the commutation relations
\eref{CG1}~\cite{Wang-2000}. The HP realization explicitly reads
\begin{eqnarray*}
  K_{+}^{\mathrm{HP}} &= a^{+}\sqrt{N^{\mathrm{SHO}}+2\eta},
  \\
  K_{-}^{\mathrm{HP}} &= \sqrt{N^{\mathrm{SHO}}+2\eta}a,
  \\
  K_{3}^{\mathrm{HP}} &= N^{\mathrm{SHO}}+\eta.
\end{eqnarray*}
where $a^{+},a$ and $N^{\mathrm{SHO}}$ are respectively the creation,
annihilation, and number operator of a single-mode electromagnetic
field and which are given by \eref{K1} and \eref{K6}. The
parameter denoted $\eta$ is the label of this representation.

The action of the generators over the Fock space is of the form
\begin{eqnarray*}
    K_{-}^{\mathrm{HP}} \ket{0} &= 0,
    \\
     K_{-}^{\mathrm{HP}} \ket{n} &= \sqrt{n(n+ 2\eta -1)}\ket{n-1},
    \\
    K_{+}^{\mathrm{HP}} \ket{n} &=  \sqrt{(2\eta + n)(n+1)}\ket{n+1},
\\
K_{3}^{\mathrm{HP}}\ket{n} &= (\eta + n)\ket{n}.
\end{eqnarray*}
which is very similar to the representation \eref{eq-144PTP}.

The principal difference between HP and the others systems(ISW, ICW, PCW
and PTP) is that for the PTP is more natural the Perelomov
coherent states that the BGCS. Here we consider the
Perelomov coherent state as a case of nonlinear coherent
states. Then, the equation that defines the nonlinear coherent
state that naturally arises for the HP system is~\cite{Wang-2000}
\begin{equation}\label{HP40}
\frac 1{N+2k}K_{-}^{\mathrm{HP}}\ket{z}=z \ket{z},
\end{equation}
where the explicit solution of \eref{HP40} is
\begin{equation}\label{HP50}
\ket{z}=(1-|z|^2)^{M/2}\sum_{n=0}^\infty {%
{M+n-1 \choose n}%
}^{1/2}z ^n\ket{n}
\end{equation}

Then, we have established that the HP also satisfy the algebraic
structure \eref{eq-150} with characteristic
functions of the form  $f^{\mathrm{HP}}(n)=n(2k+n-1)$ and
$h^{\mathrm{HP}}(N^{\mathrm{HP}})= 1/(N+2k)$ where $N^{\mathrm{HP}}=N^{\mathrm{SHO}} $. It is
possible then, to adapt the KHQA for the case of the HP with a
clearly established halting criterion.

\subsection{Laguerre oscillator}
Finally in this subsection we show that the named Laguerre
oscillator also satisfy the algebraic structure which is given by
\eref{eq-150} and for then it is possible with such system to
adapt the KHQA. The relevant formalism is the following.

We consider a Hilbert space whose elements are generalized
Laguerre functions. By constructing raising and lowering operators
acting on these states one can obtain an explicit realization of
the Hamiltonian which is defined to be diagonal in this Hilbert
space. The obtained system such as is defined by the Hamiltonian
is called Laguerre oscillator.

Now, as is well known, the Laguerre polynomials are defined  as
\begin{equation}\label{Lague00}
    L_{n}^{\alpha}(x)=\frac{1}{n!}
    e^{x}x^{-\alpha}\frac{\rmd^{n}}{\rmd x^{n}}(e^{-x}x^{\alpha +n}),
\end{equation}
and the generalized Laguerre functions are of the form
 \begin{equation}\label{Lague05}
   \psi_{n}^{\alpha}(x)=\sqrt{\frac{n!x^{\alpha
   +1}e^{-x}}{(n+\alpha)!}}L_{n}^{\alpha}(x).
\end{equation}
Now, we can define the raising operator denoted simply  $K_{+}$
and the lowering operator $K_{-}$ for the generalized Laguerre
functions:
\begin{eqnarray*}
    K_{+}\psi_{n}^{\alpha}(x) &= \left [ -x\frac{\rmd}{\rmd x}-\frac{2n+\alpha+1-x}{2} \right ]\psi_{n}^{\alpha}(x)
\\
&= -\sqrt{(n+1)(n+\alpha+1)}\psi_{n+1}^{\alpha}(x),
\\
\\
  K_{-}\psi_{n}^{\alpha}(x)&= \left [ x\frac{\rmd}{\rmd x}-\frac{2n+\alpha+1-x}{2} \right ] \psi_{n}^{\alpha}(x)
\\
&= -\sqrt{n(n+\alpha)}\psi_{n-1}^{\alpha}(x).
\end{eqnarray*}
The generalized Laguerre functions are the base of a Hilbert
space that has the structure of Fock space and at consequence
\begin{equation}\label{Lague30}
  \psi_{n}^{\alpha}(x)=\frac{1}{\sqrt{n!(\alpha
  +1)_{n}}}(K_{+})^{n}\psi_{0}^{\alpha}(x).
\end{equation}
The commutator between the ladder operators of the Laguerre
oscillator is given by
  \begin{equation}\label{Lague40}
    [K_{-},K_{+}]\psi_{n}^{\alpha}(x)=(2n+\alpha
    +1)\psi_{n}^{\alpha}(x),
\end{equation}
and then, we can define the operator denoted simply  $K_{3}$, as
\begin{equation}\label{Lague50}
    K_{3}\psi_{n}^{\alpha}(x)=\frac{1}{2}(2n+\alpha
    +1)\psi_{n}^{\alpha}(x).
\end{equation}
The commutation relations for the three operators of the Laguerre
oscillator are
 \begin{equation}\label{Lague60}
    [K_{-},K_{+}]=2K_{3},  [K_{3},K_{+}]=K_{+},
    [K_{3},K_{-}]=-K_{-},
\end{equation}
and we conclude that the Laguerre oscillator realizes a infinite-dimensional UIR of $su(1,1)$.

The Hamiltonian for the Laguerre oscillator is
\begin{equation}\label{Lague70}
    H\psi_{n}^{\alpha}(x)=K_{+}K_{-}\psi_{n}^{\alpha}(x)=e_{n}\psi_{n}^{\alpha}(x)=n(n+\alpha)\psi_{n}^{\alpha}(x),
\end{equation}
and the BGCS are defined as is usual, it is to say
\begin{equation}\label{Lague80}
    K_{-}\ket{z}=z\ket{z},
\end{equation}
where the solution of \eref{Lague80} is again the well know form
\begin{equation}\label{Lague90}
    \ket{z}=\frac{|z|^{\alpha/2}}{\sqrt{I_{\alpha}(2|z|)}}\sum_{n=0}^{\infty}\frac{z^{n}}{\sqrt{n!(n+\alpha)!}}\ket{n},
\end{equation}
where
\begin{equation}\label{Lague100}
    \ket{n}= \psi_{n}^{\alpha}(x).
\end{equation}

Then, we have proved that the Laguerre oscillator also satisfy the
algebraic structure of \eref{eq-150} with
characteristic functions of the form $f(n)=n(n+\alpha)$ y
$h(n)=1$. All this indicates that it is possible to adapt the KHQA
to case of the Laguerre oscillator.

We can observe that the Laguerre oscillator contains as particular
cases the systems ISW, ICW, PCW and PTP for different values of the
parameter $\alpha$. From the other side, it is maybe possible to
have a realization of the Laguerre oscillator and its customs
within the field of quantum optics.

\section{Conclusions}

\begin{itemize}
\item We have identified from an algebraic point of view the conditions to make adaptations of KHQA: a non-compact Lie algebra of low dimension that admits infinite-dimensional irreducible representations with naturally defined ladder operators and generalized coherent states. With base in this result, we made an adaptation of  KHQA over the algebra $\mathfrak{su}(1,1)$ due to this algebra satisfies these conditions and due to this algebra is the dynamical algebra associated to many quantum systems. 

\item Hilbert's tenth problem is a semi-computable problem by a TM in the sense that if the Diophantine equation \eref{eq-20} has solution, an exhaustive search on $k$-tuplas of non-negative integers would find it, but if \eref{eq-20} does not have solution this search would not finish. In this sense, it is possible to be interpreted ingenuously that KHQA and our adaptation over the algebra $\mathfrak{su}(1,1)$ carry out an infinite search (in a finite time) on every $k$-tuplas of non-negative integers. However, KHQA and our adaptation do not make an infinite search, due to although Hilbert's tenth problem is TM incomputable, this is a finitely refutable problem \cite{Calude-2002}. That is to say, it is only necessary to make the search on a finite set of non-negative integers, to determine if \eref{eq-20} has a solution o not, although of course, this finite set is TM incomputable.

\item A very common misunderstanding in technical literature is not to make distinction between the terms `quantum computation' and `standard quantum computation' (e.g. \cite{Cooper-2003, Braverman-Cook-2005}). Due to this misunderstanding  and due to equivalence in computability terms, between the standard quantum computation and TM computability established by David Deutsch \cite{Deutsch-1985}\footnote{In strict sense there is a type of \emph{weak hypercomputation} in standard quantum computation: the generation of truly random numbers \cite{Deutsch-1985}. Nevertheless, is not clear how using this property to solve to a TM incomputable problem \cite{Ord-Kieu-2004}.}, the hypercomputation possibility based on quantum computation is rejected. Nevertheless this situation is erroneous as it demonstrates by the theoretical existence of KHQA and our adaptation over the algebra $\mathfrak{su}(1,1)$.

\item Other common misunderstanding is not to make distinction between quantum adiabatic computation on finite and infinite-dimensional Hilbert spaces. For example, there is a recent proof that quantum adiabatic computation is equivalent to standard quantum computation \cite{Aharonov-van--Dam-Kempe-Landau-Lloyd-Regev-2004}, however this proof generates no contradiction with KHQA or with our adaptation over the algebra $\mathfrak{su}(1,1)$, due to such a proof of equivalence is only valid for quantum adiabatic computation on finite-dimensional Hilbert spaces. 

\item With base on our adaptation of KHQA over the algebra $\mathfrak{su}(1,1)$, we had presented a plausible realizations within the field of condensed matter physics and quantum optics. Although Kieu has refuted successful some critics of his algorithm (see section: \emph{Notes addes} of \cite{Kieu-2005}), there is an important observation with respect to its possible implementation that has not been solved yet, in Kieu's words: ``\emph{\dots there have been some concerns (this pointed has been raised on separate occasions by Martin Davis (2003), Stephen van Enk (2004) and Andrew Hodges (2004)) that infinite precision is still required in physically setting up the various integers parameters in the time-dependent quantum Hamiltonians. While the issue deserves further investigations as surely any systematic errors in the Hamiltonians would be fatal, we still are not convinced that such integer parameters cannot be satisfactorily set up. In particular, we would like to understand the effects of statistical (as opposed to systematic) errors on the statistical behaviour of the spectrum of our adiabatic Hamiltonians''} \cite[p. 180]{Kieu-2005a}. This observation is valid for our plausible realizations too, however we agree it is necessary further investigations to establish if it is possible or not to implement KHQA or our adaptation over the algebra $\mathfrak{su}(1,1)$.

\end{itemize}

\ack
We thank to Professor J. P. Antoine for help us with some definitions. We thank to Professor Tien~D.~Kieu for
helpful discussions and feedback. One of us (A. S.) would likes to acknowledgment the kind hospitality during his visit to Prof. Kieu at the CAOUS at Swinburne University of Technology. We are also thankful to some
anonymous referees for their accurate observations and suggestions
to preliminary versions of this article. This research was
supported by COLCIENCIAS-EAFIT (grant \# 1216-05-13576).

\appendix
\section*{Appendix}
\setcounter{section}{1}

In this appendix we present the procedure to obtain the explicit
form of the coherent states denoted  $\ket{z}^\mathrm{S}$, which is
given by \eref{CG9} and which was used to obtain the explicit
forms for the all particular coherent states, both of the
Barut-Girardello as the Perelomov type, that were used in this
work. Since that the coherent state $\ket{z}^\mathrm{S}$ belongs to
the Fock space $\mathfrak{F}^{\mathrm{S}}$ \eref{CG0}, we can to write the
coherent state as an linear combination
\begin{equation}\label{ap-10}
   \ket{z}^{\mathrm{S}} =\sum_{n=0}^{\infty}C_{n}(z)\ket{n}.
 \end{equation}
The substitution of \eref{ap-10} on \eref{CG8} and using
\eref{CG2} and \eref{CG5} generates the following recurrence
equation for the coefficients $C_{n}(z)$
\begin{equation}\label{ap-20}
    C_{n+1}(z)h^{\mathrm{S}}(n)\sqrt{f^{\mathrm{S}}(n+1)}=z C_{n}(z).
\end{equation}
The solution of \eref{ap-20} is
\begin{equation}\label{ap-30}
  C_{n}(z)=C_{0}(z)\frac{z^{n}}{\left( \prod_{j=0}^{n-1}h^{\mathrm{S}}(j) \right ) \left( \sqrt{f^{\mathrm{S}}(n)!} \right )}.
\end{equation}
To obtain the coefficient $C_{0}(z)$ we apply the condition of
normalization of the coherent state
\begin{equation}\label{ap-40}
    ^{\mathrm{S}}\braket{z}{z}^{\mathrm{S}} = 1 = \sum_{n=0}^{\infty} C_{0}(z)^{2}\frac{|z|^{2n}}{\left( \prod_{j=0}^{n-1}h^{\mathrm{S}}(j)\right )^{2} \left (f^{\mathrm{S}}(n)! \right )}.
\end{equation}
From \eref{ap-40} we obtain that
\begin{equation}\label{ap-50}
C_{0}(z)= \left ( \sum_{m=0}^{\infty}
  \frac{|z|^{2m}}{\left ( \prod_{j=0}^{m-1}h^{\mathrm{S}}(j)\right )^{2} \left ( f^{\mathrm{S}}(m)! \right )} \right )^{-1/2}  .
\end{equation}
Finally the substitution of \eref{ap-50} on \eref{ap-30} and
then on \eref{ap-10}, gives the following explicit form for the
$\mathfrak{su}(1,1)$ non-linear coherent states which is given by \eref{CG9}.

\section*{References}
\bibliographystyle{unsrt}
\bibliography{/home/asicard/bibtex/asicard}

\end{document}